\documentclass[twocolumn,prd,showpacs]{revtex4}

\usepackage{graphicx}
\usepackage{dcolumn}
\usepackage{bm}

\newcommand {\beq} {\begin{equation}}
\newcommand {\eeq} {\end{equation}}
\newcommand {\beqa}{\begin{eqnarray}}
\newcommand {\eeqa}{\end{eqnarray}}
\newcommand {\del} {\partial}

\newcommand {\Tr}{\mbox{Tr\,}}

\newcommand {\dd}{\mbox{d}}
\newcommand {\ee}{\mbox{e}}

\newcommand {\defeq}{\stackrel{\rm def}{=}}

\begin{document}

\preprint{NBI--HE--01--09}

\title{A new approach to the complex-action problem\\
and its application to a nonperturbative study of superstring theory}

\author{K.N.\ Anagnostopoulos}
\email{konstant@physics.uoc.gr}
\affiliation{%
Department of Physics, University of Crete,
P.O. Box 2208, GR-71003 Heraklion, Greece
}
\author{J.\ Nishimura}
\email{nisimura@eken.phys.nagoya-u.ac.jp}
\affiliation{%
The Niels Bohr Institute,
Blegdamsvej 17, DK-2100 Copenhagen \O, Denmark, and\\
Department of Physics, 
Nagoya University, Furo-cho, Chikusa-ku, Nagoya 464-8602, Japan 
}%

\date{\today}

\begin{abstract}
Monte Carlo simulations of a system whose action has an imaginary
part are considered to be extremely difficult.
We propose a new approach to this `complex-action problem',
which utilizes a factorization property of distribution functions.
The basic idea is quite general,
and it removes the so-called overlap problem completely.
Here we apply the method to a nonperturbative study of superstring theory
using its matrix formulation.
In this particular example, the distribution function turns out
to be positive definite, which allows us to reduce the problem even further.
Our numerical results suggest an intuitive explanation for the
dynamical generation of 4d space-time.

%
%
\end{abstract}


\maketitle

\section{\label{sec:intro}Introduction}

It occurs in many interesting systems 
ranging from condensed matter physics to high-energy physics
that their action has an imaginary part.
Some examples for instance in high-energy physics
are the finite density QCD, Chern-Simons theories,
systems with topological terms (like the $\theta$-term in QCD),
and systems with chiral fermions.
While this is not a conceptual problem, it poses a technical problem
when one attempts to study these systems
by Monte Carlo simulations, which
would otherwise provide a powerful tool to understand their properties 
from first principles
(see Refs.\ \cite{signproblem,Fodor:2001au,meron} for recent works).

In this Letter we propose a new approach to this `complex-action problem'.
Suppose we want to obtain an expectation value of some observable.
Then, as a more fundamental object,
we consider the distribution function associated with that observable.
In general the distribution function has
a factorization property, which relates it
to the distribution function 
associated with the same observable but calculated
{\em omitting} the imaginary part of the action.
The effect of the imaginary part is represented by
a correction factor which can be obtained by a constrained Monte Carlo 
simulation.
One of the virtues of this method
is that it removes the so-called overlap
problem completely.
This problem comes from the fact that the two distribution functions
--- one for the full model and the other for the model omitting the
imaginary part --- have little overlap in general.
The method avoids this problem by `forcing' 
the simulation to sample the
important region for the full model.

The determination of the correction factor becomes increasingly
difficult as the system size increases.
In this sense, our approach does not solve the complex action problem
completely. This should be contrasted to 
the meron-cluster algorithm \cite{meron},
with which one can study a special class of complex-action systems
by computer efforts increasing at most by some power of the system
size. 
The factorization method eliminates the overlap problem,
which composes some portion of the complex action problem,
but not the whole.
However, the resolution of the overlap problem is 
in fact a substantial progress.
For instance, Refs.\ \cite{Fodor:2001au} developed a new 
method to weaken the same problem in finite density QCD,
and the critical point was successfully identified.
Therefore we expect that the {\em complete} resolution 
of the overlap problem 
allows us to address various interesting questions
related to complex-action systems
with the present computer resources.
Since our method is based on the general property of 
distribution
functions, it can be applied to {\em any} complex-action systems.

In this article we are concerned with a nonperturbative
study of superstring theory using its matrix formulation \cite{IKKT}.
Eventually we would like to examine the possibility
that our 4-dimensional space time appears dynamically in
10-dimensional string theory \cite{AIKKT,HNT,NV,branched,Burda:2000mn,4dSSB}.
Monte Carlo simulation of the matrix model suffers from the complex
action problem, and there are evidences that the imaginary part of the
action plays a crucial role in 
the dynamical reduction of the space-time dimensionality \cite{NV}.
We will discuss how we can study such an issue by Monte Carlo 
simulation using the new approach.
%


\section{\label{sec:model}
The superstring matrix model}

As a nonperturbative definition of type IIB superstring
theory in 10 dimensions,
Ishibashi, Kawai, Kitazawa and Tsuchiya \cite{IKKT} proposed
a matrix model, which can be formally obtained
by the zero-volume limit of $D=10$, ${\cal N}=1$, 
pure super Yang-Mills theory.
The partition function of the type IIB matrix model
(and its obvious generalizations to $D=4,6$) can be written as
\beq
Z =  \int \dd A ~ \ee^{-S_{\rm b} }~Z_{\rm f} [A]   \ ,
\label{original_model}
\eeq
where $A_{\mu}$ ($\mu = 1,\cdots , D$) are 
$D$ bosonic $N \times N$ traceless hermitian matrices,
and 
$S_{\rm b}= - \frac{1}{4g^2}\Tr([A_{\mu},A_{\nu}]^2)$
is the bosonic part of the action.
The factor $Z_{\rm f} [A]$ represents the quantity
obtained by integration over the fermionic matrices,
and its explicit form is given for example in 
Refs.\ \cite{NV,AABHN}.
The convergence of the integral (\ref{original_model})
for arbitrary $N \ge 2$
was first conjectured \cite{KNS} and proved recently \cite{AW}.
The only parameter $g$ in the model can be absorbed by
rescaling $A_\mu \mapsto \sqrt{g} A_\mu$, which means
that $g$ is merely a scale parameter rather than a coupling constant.
Therefore, one can determine the $g$ dependence of any quantities 
on dimensional grounds \cite{endnote0}. 
Throughout this paper, we make our statements
in such a way that they do not depend on the choice of $g$.

In this model space-time is represented by $A_{\mu}$,
and hence treated dynamically \cite{AIKKT}.
It is Euclidean as a result of the Wick rotation,
which is always necessary in path integral formalisms.
Its dimensionality is dynamically determined and can be
probed by the moment of inertia tensor defined by \cite{HNT}
\beq
T_{\mu\nu} = \frac{1}{N} \Tr (A_\mu A_\nu) \ .
\eeq
Since $T_{\mu\nu}$ is a $D \times D$ real symmetric matrix,
it has $D$ real eigenvalues corresponding to the 
principal moments of inertia,
which we denote as $\lambda_i$
($i=1, \cdots , D$) with the ordering
\beq
\lambda_1 > \lambda _2 > \cdots > \lambda _{D} >  0 \ .
\label{ordering}
\eeq
Let us define the VEV $\langle {\cal O} \rangle$ with respect
to the partition function (\ref{original_model}).
If we find that
$\langle \lambda_i \rangle$ with $i = 1,\cdots , d$ is much larger
than the others, we may conclude 
that the dimensionality of the dynamical space-time is $d$ .

\section{\label{sec:problem}The complex action problem}

The fermion integral $Z_{\rm f} [A]$ 
in the partition function (\ref{original_model})
is complex in general for $D=10$, $N\ge 4$ 
and for $D=6$, $N\ge 3$ \cite{NV}.
Let us restrict ourselves to these cases in what follows.
Parameterizing the fermion integral as
$Z_{\rm f} [A] = \exp (\Gamma_{\rm R} + i \Gamma )$,
the partition function (\ref{original_model}) may be written as
\beq
Z =  \int \dd A ~ \ee^{-S_0}~\ee^{i \Gamma}  \ ,
\label{fullZ}
\eeq
where $S_0 = S_{\rm b} - \Gamma_{\rm R}$ is real. 
According to the standard reweighting method,
one evaluates the VEV $\langle \lambda_i \rangle$ as
\beq
\left\langle \lambda_i \right\rangle
= \frac{\left\langle \lambda_i \, \ee ^{i \Gamma }
\right\rangle _{0}}
{\left\langle \ee ^{i \Gamma }
\right\rangle _{0}}  \ ,
\label{VEV}
\eeq
where the symbol $\langle \ \cdot \ \rangle _{0}$ denotes 
a VEV with respect to the partition function
\beq
Z_0  =  \int \dd A ~ \ee^{-S_0 }  \ .
\label{model_0}
\eeq
The VEV $\langle \, \cdot \, \rangle _{0}$ can be evaluated by
standard Monte Carlo simulations.
However, 
$\left\langle \ee ^{i \Gamma }\right\rangle _{0}$
is nothing but the ratio of the two partition functions $Z_0/Z$,
and therefore it behaves as $\ee ^{- N^2 \Delta F}$ at large $N$, where
$\Delta F >0$ is the difference of the free energy density of the
corresponding two systems.
This enormous cancellation 
(note that $| \ee ^{i \Gamma }| = 1$ for each configuration)
is caused by 
the fluctuation of the phase $\Gamma $, 
which grows linearly with the number of fermionic
degrees of freedom, which is of O($N^2$).
As a result
the number of configurations required to obtain the VEV
$\left\langle \ee ^{i \Gamma }\right\rangle _{0}$
with sufficient accuracy grows as $\ee ^{{\rm const.} N^2}$.
The same is true for the numerator 
$\left\langle \lambda_i \, \ee ^{i \Gamma } \right\rangle _{0}$
in (\ref{VEV}).
This is the notorious `complex action problem'
(or rather the `sign problem', as we see below),
which occurs also in many other interesting systems.

In fact we may simplify the expression (\ref{VEV}) by
using a symmetry.
We note that under parity transformation :
\beq
\left\{ \begin{array}{ll}
A^P _1  = - A_1   & ~  \\
A^P_i = A_i  & \mbox{for~}2 \le i \le D \ ,
\end{array} 
\right. 
\label{paritytr}
\eeq
the fermion integral $Z_{\rm f} [A]$ 
becomes complex conjugate \cite{NV}, while
the bosonic action $S_{\rm b}$ is invariant.
Since the observable $\lambda_i$ is also invariant,
we can rewrite (\ref{VEV}) as
\beq
\left\langle \lambda_i \right\rangle
= \frac{\left\langle \lambda_i \, \cos  \Gamma 
\right\rangle _{0}}
{\left\langle \cos  \Gamma 
\right\rangle _{0}}  \ .
\label{VEV2}
\eeq
Note, however, that the problem still remains,
since $\cos  \Gamma $ flips its sign violently
as a function of $A_\mu$.

\section{\label{sec:method}The new method}

\subsection{\label{sec:factorization}The factorization property
of distributions}

The model (\ref{model_0}) omitting the phase $\Gamma$
was studied up to $N=768$ and $N=512$ for $D=6$ and $D=10$ 
respectively using the low-energy effective theory \cite{AIKKT}.
There it was found that $\langle \lambda_i \rangle_0/ (g N^{1/2})$ approaches
a universal constant independent of $i$ as $N$ increases.
This means that 
the dynamical space-time becomes isotropic in $D$-dimensions at
$N=\infty$, and hence the absence of SSB of SO($D$) symmetry,
{\em if one omits the phase} $\Gamma$.

We normalize the principal moments of inertia $\lambda_i$ as
\beq
\tilde{\lambda} _i 
\defeq \frac{\lambda _i}{\langle \lambda_i \rangle_0} \ .
\label{normalizeL}
\eeq
Then the deviation of $\langle \tilde{\lambda} _i \rangle$ from 1
represents the effects of the phase.
The relevant question is whether the deviation
depends on $i$ at large $N$.
In order to obtain the expectation value 
$\langle \tilde{\lambda} _i \rangle$,
we consider the distribution associated with the observable
$\tilde{\lambda} _i$ :
\beq
\rho_i (x) 
\defeq \langle \delta (x - \tilde{\lambda} _i) \rangle \ .
\label{defrho}
\eeq
As an important property of the distribution $\rho _i (x)$,
it factorizes as
\beq
\rho_i (x)  = \frac{1}{C} \, \rho ^{(0)} _i (x) \, 
w_i (x)  \ ,
\label{key-identity}
\eeq
where $C$ is a normalization constant given by 
\beq
C \defeq \langle \ee ^{i \Gamma}  \rangle _0= 
\langle \cos  \Gamma  \rangle_0 \ .
\eeq
The real positive function $\rho ^{(0)}_i (x)$ is defined by
\beq
\rho ^{(0)}_i (x) \defeq \langle \delta (x - \tilde{\lambda}_i) 
\rangle_0 \ ,
\eeq
which is nothing but the distribution of $\tilde{\lambda}_i$
in the model (\ref{model_0}) without $\Gamma$.
The function $\rho _i ^{(0)} (x)$ is peaked at $x=1$ due to the 
chosen normalization (\ref{normalizeL}).
The function $w_i (x)$ in
(\ref{key-identity}) can be regarded as
the correction factor representing the effect of $\Gamma$,
and it is given explicitly as 
\beq
w_i (x) \defeq
\langle \ee ^{i \Gamma}  \rangle _ {i,x} =
\langle \cos  \Gamma  \rangle _ {i,x} \ ,
\eeq
where 
the symbol $\langle \ \cdot \ \rangle _ {i,x}$
denotes a VEV with respect to a yet another partition function
\beq
Z_ {i,x} = \int \dd A \, 
\ee ^{-S_0 } \, \delta (x - \tilde{\lambda}_i) \ .
\label{cnstr_part}
\eeq
Given all these definitions, it is straightforward to prove
the relation (\ref{key-identity}).

\subsection{\label{sec:Monte}Monte Carlo evaluation of 
$\rho^{(0)}_i(x)$ and $w_i(x)$} 

In order to obtain the function $w_i (x)$, we have to simulate
(\ref{cnstr_part}).
In practice we simulate instead the system
\beq
Z_{i, V}
 =  \int \dd A ~ \ee^{-S_0 }~ \ee ^{- V(\lambda_i) }  \ ,
\label{part_pot}
\eeq
where $V(\lambda_i)$ is some potential introduced
only for the $i$-th principal moment
of inertia. 
The explicit form of the potential we used
in the study is
$V(z) = \frac{1}{2} \gamma (z -\xi)^2 $, where 
$\gamma$ and $\xi$ are real parameters.
The results are insensitive to $\gamma$ as far as it is sufficiently
large and we took $\gamma = 1000.0$.
Let us denote the VEV associated with 
the partition function (\ref{part_pot})
as $\langle {\cal O}\rangle _{i,V}$.
Then the expectation value $\langle \cos \Gamma \rangle _{i,V}$
provides the value of $w_i (x)$ at
$x = \langle \lambda _ i \rangle _{i,V}$.

The function $\rho ^{(0)}_i (x)$ can be obtained 
from the same simulation (\ref{part_pot}).
Note that the distribution function for 
$\tilde{\lambda}_i$ in the system (\ref{part_pot}) is given by
\beq
\rho _{i,V} (x) \defeq 
\langle \delta (x - \tilde{\lambda}_i) \rangle_{i,V}
\propto 
\rho ^{(0)}_{i} (x) \ee ^{-V(\langle \lambda_i \rangle_0 x)} \ .
\eeq
The position of the peak $x_{\rm p}$ is given by the solution to 
\beq
0 = \frac{\del}{\del x} \ln \rho _{i,V} (x) 
 = f^{(0)}_i (x)
- \langle \lambda_i \rangle_0 V'(\langle \lambda_i \rangle_0 x) \ ,
\eeq
where we have defined
\beq
f^{(0)}_i (x) \defeq 
\frac{\del}{\del x} \ln  \rho ^{(0)}_{i} (x) \ .
\eeq
This implies that
$\langle \lambda_i \rangle_0 V'(\langle \lambda_i \rangle_0 x_{\rm
p})$ gives the value of $f^{(0)}_i (x)$ at $x = x_{\rm p}$.
Since we take $\gamma$ sufficiently large, the distribution
$\rho _{i,V} (x)$ has a sharp peak, and we can safely replace
the position of the peak $x_{\rm p}$ by the expectation value
$\langle \tilde{\lambda}_i \rangle_{i,V}$.
Once we obtain $f_i ^{(0)}(x)$, we can obtain $\rho ^{(0)}_{i} (x)$
by
\beq
\rho ^{(0)}_{i} (x) = \exp 
\left[\int _0 ^x  \dd z  \,
f_i ^{(0)}(z) + {\rm const.} \right] , 
\eeq
where the integration constant can be determined
by the normalization of $\rho^{(0)}_{i} (x)$.

\begin{figure*}
\includegraphics{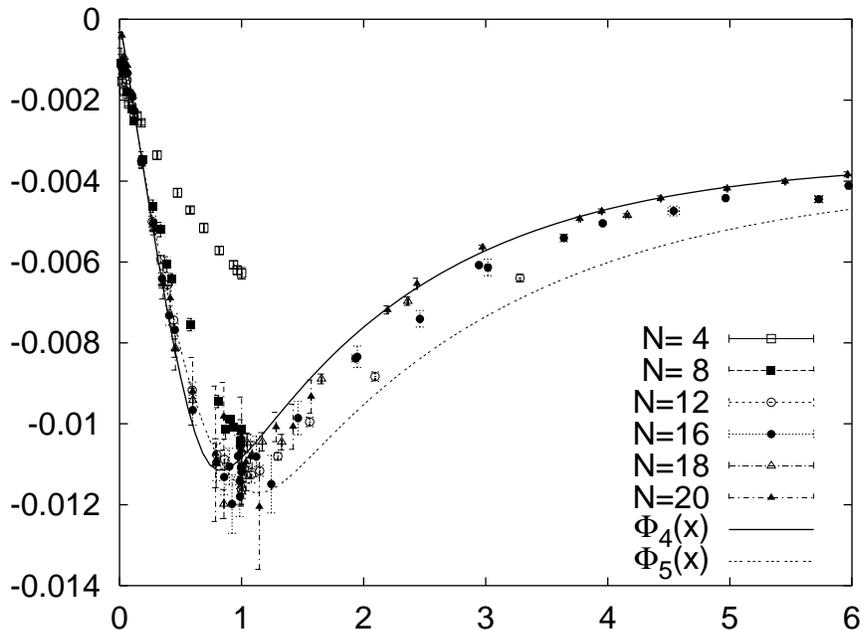}
\caption{\label{fig:scaling} 
The function $\frac{1}{N^2} \ln w_4 (x)$
is plotted 
for $N=$12,16,18,20.
For $x<1$ we also plot data for $N=4,8$ to clarify the convergence.
We extract the scaling function $\Phi_4(x)$ by fitting the data
to some analytic function, which is represented by
the solid line.
The dashed line represents $\Phi_5 (x)$, which is obtained
similarly from the scaling behavior of $\frac{1}{N^2} \ln w_5 (x)$.
}
\end{figure*}

\subsection{\label{sec:overlap}Resolution of the overlap problem}

From $\rho^{(0)}_i(x)$ and $w_i(x)$, 
we may obtain the VEV $\langle\tilde{\lambda}_i \rangle$ by
\beq
\langle \tilde{\lambda}_i \rangle
= \int _0^{\infty} \dd x \, x \, \rho_i(x) 
= \frac{\int _0^{\infty} \dd x \, x \, \rho^{(0)}_i(x) \,  w_i(x) }
{\int _0^{\infty} \dd x  \, \rho^{(0)}_i(x) \, w_i(x) }
 \ .
\label{naive}
\eeq
Actually this simply amounts to using the reweighting formula
(\ref{VEV2}) but calculating the VEVs on the r.h.s.\ by
\beqa
\label{similar0}
\langle \tilde{\lambda}_i \cos \Gamma \rangle_0
&=& \int_0^{\infty} \dd x \, x \, \rho^{(0)}_i(x) \, w_i(x) \\
\langle  \cos \Gamma \rangle_0
&=& \int_0^{\infty} \dd x \, \rho^{(0)}_i(x) \, w_i(x) \ .
\label{similar}
\eeqa
This reveals one of the virtues of our approach
as compared with the standard reweighting method 
using the formula (\ref{VEV2}) directly.
Suppose we are to obtain the l.h.s.\ of (\ref{similar0})
and (\ref{similar})
by simulating the system (\ref{model_0}).
Then for most of the time, $\tilde{\lambda}_i$ takes the value
at the peak of $\rho^{(0)}_i(x)$.
However, in order to obtain the VEVs accurately we have to 
sample configurations whose $\tilde{\lambda}_i$ takes a value
where $|\rho^{(0)}_i(x) w_i(x)|$ becomes large.
In general the overlap of the two functions becomes exponentially
small with the system size,
and this makes the important sampling ineffective.
Therefore, this `overlap problem' composes some portion
of the complex-action problem.
The new approach eliminates this problem by `forcing' the
simulation to sample the important region.

\begin{figure*}
\includegraphics{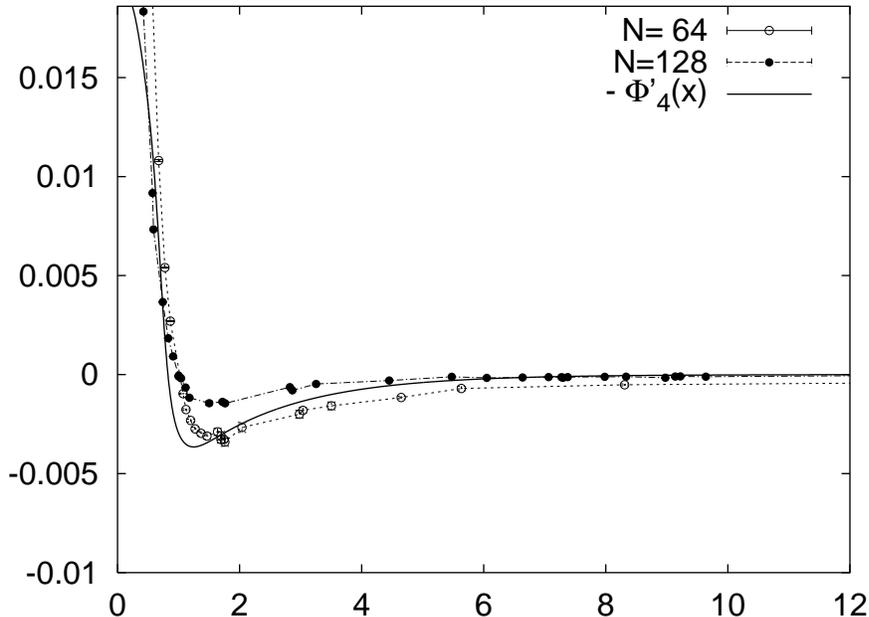}
\caption{\label{fig:xMUplotA4} 
The function $\frac{1}{N^2} f ^{(0)} _4 (x)$
is plotted 
for $N=64,128$.
The solid line represents $- \Phi_4 '(x)$, which we calculate
from the scaling function $\Phi _4 (x)$ extracted
in Fig.\ \ref{fig:scaling}.
}
\end{figure*}

\subsection{\label{sec:scaling}Further improvement in the case
$w_i(x)>0$}

So far, we have been discussing the general properties of the new
method. 
In the case at hand, we can actually further reduce the problem
by using the fact that 
the correction factors $w_i(x)$ are actually {\em positive definite},
and so is the full distribution function $\rho _i (x)$.
(Note that this is not guaranteed {\em in general}.)
This allows us to obtain the VEV $\langle \tilde{\lambda}_i
\rangle$ by minimizing the `free energy density' 
$F_i (x) = - \frac{1}{N^2}\log \rho _i (x)$,
instead of using
(\ref{naive}).
For that we simply need to solve
$F' _i (x) = 0$, which is equivalent to 
\beq
\frac{1}{N^2} f^{(0)}_i (x) 
= - \frac{d}{dx} \left[ \frac{1}{N^2} \ln w_i (x) \right] \ .
\label{diff_eq}
\eeq
The function in the bracket $[ \ \cdot \ ]$
is expected to approach a well-defined function
as $N$ increases :
\beq
\frac{1}{N^2} \ln w_i (x)  \rightarrow \Phi_i(x) \ .
\label{fscaling}
\eeq
Let us note that
$w_i(x)$ is nothing but the expectation value 
of $\ee^{i \Gamma}$ in the system (\ref{cnstr_part}).
According to the argument below (\ref{model_0}),
$w_i(x)$ for fixed $x$ decreases as 
$\ee ^{- \alpha N^2}$ at large $N$.
The constant $\alpha$ may depend on $x$, hence the assertion.
Indeed our numerical results in Fig.\ \ref{fig:scaling}
(although the achieved values of $N$ 
are not very large) seem to support this argument.
Once we extract the scaling function $\Phi_i(x)$,
we may use it instead of $\frac{1}{N^2} \ln w_i (x)$ 
in (\ref{diff_eq}) for larger $N$.
Thus we are able to obtain
the VEV $\langle \tilde{\lambda}_i \rangle$ for much larger
$N$ than those allowing the direct Monte Carlo evaluation of 
the correction factor $w_i(x)$.

The positive definiteness
of $w_i(x)$ is crucial for such an extrapolation technique
to work.
If we were to calculate the VEV $\langle \tilde{\lambda}_i \rangle$
by (\ref{naive}), we would need to calculate
the correction factor for larger $N$ by
$w_i(x)=\ee^{N^2 \Phi_i(x)}$,
where the multiplication by $N^2$ and the exponentiation would
magnify the errors in $\Phi_i(x)$ considerably.
This does {\em not} occur when we obtain 
the VEV $\langle \tilde{\lambda}_i \rangle$ by solving (\ref{diff_eq}).

\section{\label{sec:results}Results}

Monte Carlo simulation of (\ref{part_pot}) can be
performed by using the algorithm developed 
for the model (\ref{model_0}) in Ref.\ \cite{AABHN}.
The required computational effort is O($N^6$).
In this work, we use instead the low-energy effective theory
proposed in Ref.\ \cite{AIKKT} and further developed in
Ref.\ \cite{branched}.
The required computational effort becomes O($N^3$).
For the definition of the low-energy effective theory
as well as all the technical details including
parameters used in the simulations, we follow 
Ref.\ \cite{branched}.
The validity of the low-energy effective theory in
studying the extent of the dynamical space
time is discussed in Ref.\ \cite{AABHN}.
We also note that the complex-action problem
survives in passing from the full theory to 
the low-energy effective theory, and hence
we expect that the effects of the phase
on the reduction of space-time dimensionality
should be visible also in the low-energy effective theory,
if it is there at all.
Here we study the $D=6$ case (instead of $D=10$,
which corresponds to the type IIB matrix model)
to decrease the computational efforts further.

In Fig.\ \ref{fig:scaling},
we plot $\frac{1}{N^2} \ln w_4 (x)$.
The correction factor $w_4 (x)$
has a minimum at $x \sim 1$
and it becomes larger for both $x<1$ and $x>1$.
This can be understood as follows.
Let us recall again that $w_i(x)$ is the expectation value 
of $\ee^{i \Gamma}$ in the system (\ref{cnstr_part}),
where $\tilde{\lambda}_i$ is constrained to a given value of $x$.
At $x = 1$, the system (\ref{cnstr_part}) is almost equivalent to 
the system (\ref{model_0}), because $\tilde{\lambda}_i$ would be
close to 1 even without the constraint.
(From this, it also follows that $w_i(1)$ takes almost the 
same value for all $i$.)
Therefore, the dominant configurations of the system (\ref{cnstr_part})
at $x=1$ is isotropic at large $N$ \cite{branched}.
On the other hand,
the dominant configurations of the system (\ref{cnstr_part})
at small $x$ are $(i-1)$-dimensional, since the constraint
forces $\tilde{\lambda}_i$ to be small, and due to the 
ordering (\ref{ordering}), all the $\tilde{\lambda}_j$ with $j\ge i$
become small.
Similarly 
the dominant configurations of the system (\ref{cnstr_part})
at large $x$ are almost $i$-dimensional, since the constraint
forces $\tilde{\lambda}_i$ to be large, and due to the 
ordering (\ref{ordering}), all the $\tilde{\lambda}_j$ with $j\le i$
become large. 
Now let us recall that the phase $\Gamma$ vanishes
when the configuration $A$ has the 
dimensionality $d \le d_{\rm cr}$, where 
$d_{\rm cr} = 4, 6$ for $D=6,10$, respectively \cite{NV}.
As a consequence, $w_4 (x)$ gets larger
in both $x<1$ and $x>1$ regimes.

As mentioned already,
Fig.\ \ref{fig:scaling} supports
the scaling behavior (\ref{fscaling})
with increasing $N$.
The scaling function $\Phi_4(x)$ can be extracted
by fitting the data to some analytic function.
We find that $\Phi_4(x)$ approaches 0 linearly as $x \rightarrow 0$,
and it approaches some negative constant exponentially
as $x \rightarrow \infty$.
We observe a similar scaling behavior for $\frac{1}{N^2} \ln w_5 (x)$.
The corresponding scaling function $\Phi_5(x)$ is plotted in
Fig.\ \ref{fig:scaling} for comparison.

\begin{figure*}
\includegraphics{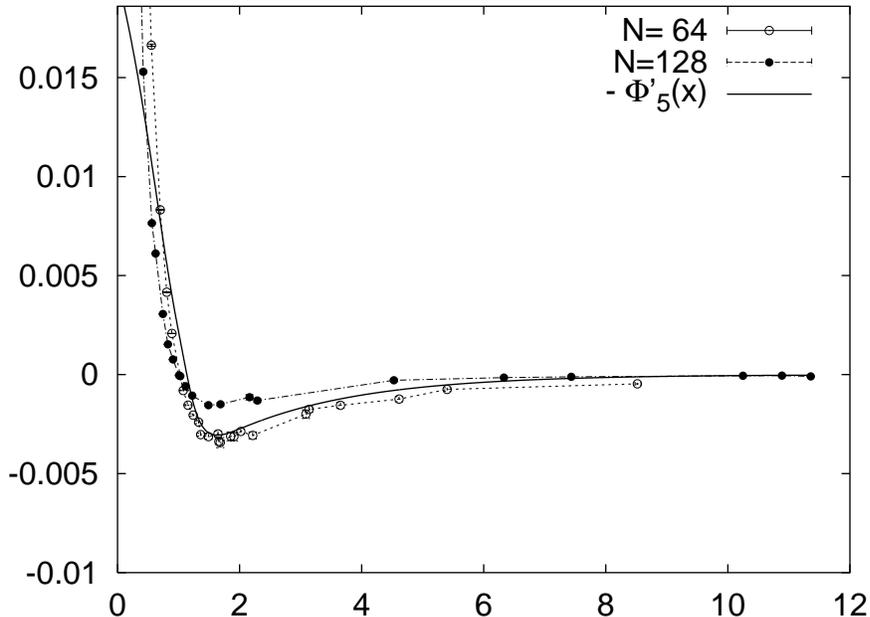}
\caption{\label{fig:xMUplotA5} 
The function $\frac{1}{N^2} f  ^{(0)} _5 (x)$
is plotted 
for $N=64,128$.
The solid line represents $- \Phi_5 '(x)$, which we calculate
from the scaling function $\Phi _5 (x)$ shown in Fig.\
\ref{fig:scaling}.
}
\end{figure*}

Fig.\ \ref{fig:xMUplotA4} represents
a graphical solution of (\ref{diff_eq}) for $i=4$.
The open and closed circles
describe the function $\frac{1}{N^2} f ^{(0)} _4 (x)$ 
for $N=64,128$ respectively.
It is positive at $x<1$ and turns negative at $x>1$,
which reflects the fact that
$\rho _i ^{(0)} (x)$ is peaked at $x=1$.
The solid line represents $- \Phi' _4 (x)$.
The intersections of the two curves provide
the solutions to (\ref{diff_eq}).
At $N=128$, we find that the distribution $\rho_4 (x)$ has
two peaks; one at $x=x_{s}<1$ and the other at $x=x_{l}>1$.
The ratio of the peak height $R = \rho_4 (x_{s})/\rho_4 (x_{l})$
can be written as 
$R = \exp \{ N^2 ({\cal A}_{s} - {\cal A}_{l})\} $, where
${\cal A}_{s}$ and ${\cal A}_{l}$ 
are the area of the regions surrounded by the
two curves.
We obtain ${\cal A}_{s} \sim 5.0 \times 10^{-4}$ and
$ {\cal A}_{l}\sim 4.5\times 10^{-3}$,
from which we conclude that the
peak at $x >1$ is dominant.
In Fig.\ \ref{fig:xMUplotA5} we show the results
of a similar analysis for $\rho_5(x)$.
We find that the distribution $\rho_5 (x)$ at $N=128$ also has
two peaks; one at $x<1$ and the other at $x>1$.
However, here we obtain ${\cal A}_{s} \sim 2.0 \times 10^{-3}$ and
$ {\cal A}_{l}\sim 3.8 \times 10^{-3}$, which are comparable.

\section{\label{sec:summary}Summary and discussions}

We have proposed a new method to study complex-action systems
by Monte Carlo simulations.
In particular we discussed how we can use the method 
to investigate the possibility 
that four-dimensional space time is dynamically generated 
in the type IIB matrix model.
The space-time dimensionality is probed by
the eigenvalues $\lambda_i$ of the moment of inertia tensor
and we study the distribution of each eigenvalue.
The distribution $\rho _i ^{(0)}(x)$ obtained without the phase $\Gamma$
has a single peak, which is located at $x=1$.
The effect of the phase $\Gamma$ on the distribution function
is represented by the multiplication
of the correction factor $w_i (x)$
as stated in (\ref{key-identity}).

Our results for the 4-th and 5-th eigenvalues ($i=4,5$)
in the $D=6$ case show that
the correction factor $w_i (x)$ strongly suppresses the peak
of $\rho _i ^{(0)}(x)$ at $x=1$ and favours {\em both} smaller $x$
and larger $x$.
As a result, we observe that the distribution $\rho_i (x)$ including the 
effects of the phase, in fact, has a {\em double peak structure}.
Moreover, the two peaks tend to move away from $x=1$ as $N$ is increased.
It is important to determine which of the two peaks becomes dominant
in the large $N$ limit.
At $N=128$, we observe that the peak at $x>1$ is dominant
for both $\rho _4 (x)$ and $\rho _5 (x)$.
We note, however, that
it is much more dominant for $\rho _4 (x)$ than for $\rho _5 (x)$.



\begin{figure*}
\includegraphics{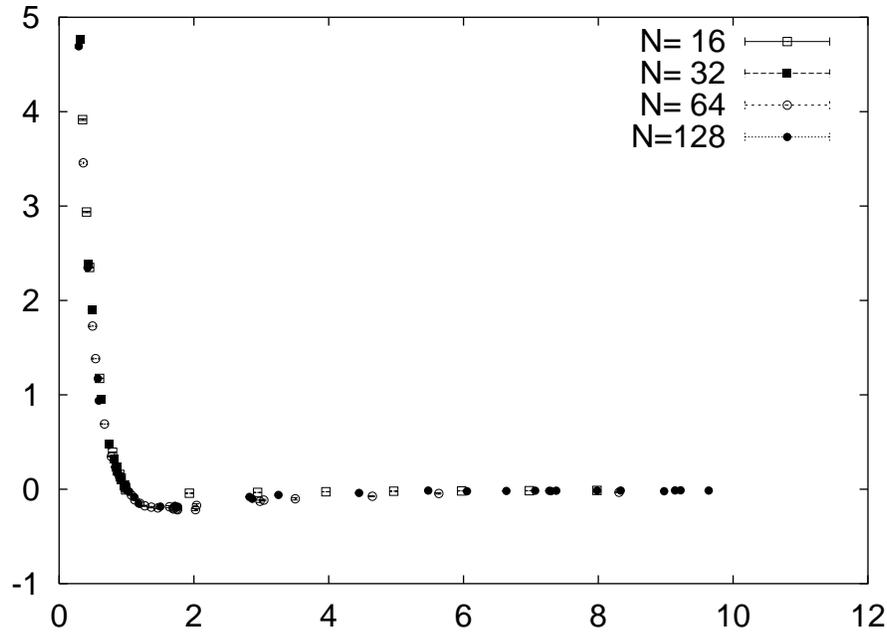}
\caption{\label{fig:f0scale}
The function $\frac{1}{N} f  ^{(0)} _4 (x)$
is plotted 
for $N=16,32,64,128$.
A clear scaling behavior is observed.
}
\end{figure*}

From Figs.\ \ref{fig:xMUplotA4} and \ref{fig:xMUplotA5},
we observe that the function $\frac{1}{N^2}f^{(0)}_i (x)$ 
changes drastically as we go from $N=64$ to $N=128$.
In fact we find that $\frac{1}{N}f^{(0)}_i (x)$ scales
(notice the normalization factor $\frac{1}{N}$),
as shown in Fig.\ \ref{fig:f0scale} for $i=4$.
The scaling region extends from 
$x\sim 1$, where $\frac{1}{N}f^{(0)}_i (x)$ crosses zero,
namely the place where $\rho ^{(0)}_i (x)$ has a peak.
A similar scaling behavior is observed for $i=5$.
This scaling behavior is understandable if we recall
that the long-distance property of the system 
is controlled by 
a branched-polymer like system \cite{AIKKT},
which is essentially a system with $N$ degrees of freedom.
If we {\em naively} extrapolate this scaling behavior of 
$\frac{1}{N}f^{(0)}_i (x)$ to larger $N$,
the l.h.s.\ of (\ref{diff_eq}) becomes
negligible. It follows that
the peak at $x<1$ eventually dominates for both $i=4,5$,
considering the asymptotic behaviors of $\Phi_i(x)$ as 
$x\rightarrow 0$ and $x\rightarrow \infty$.
This means that the space-time dimensionality becomes $d \le 3$.
However, 
it is well-known that the Hausdorff dimension 
of a branched polymer is 
$d_{\rm H}=4$, which implies
that such a system
is not easy to collapse into a configuration with dimensions $\le 3$.
The consequence would be that
$\rho^{(0)} _4 (x)$ is much more suppressed in the small $x$ regime
than $\rho^{(0)} _5 (x)$ at large $N$.
We consider that
this prevents the peak at $x<1$ from dominating
for $\rho_4 (x)$, and as a result we obtain 4d space-time.
Since the above argument is based only on the scaling behaviors
and the branched polymer description,
it is expected to be valid also in the $D=10$ case.
(While this paper was being revised, 
an analytic evidence for the dominance of 4d space-time 
was also reported 
\cite{NS}.)





Our new approach to complex-action systems
is based on 
the factorization property (\ref{key-identity})
of distribution functions,
which is quite general.
As we discussed in Section \ref{sec:overlap},
it resolves the overlap problem completely.
In a separate paper we will report on a test
of the new method
in a Random Matrix Theory for finite density QCD, 
where exact results in the thermodynamic limit are
successfully obtained \cite{prep}.
We hope that the `factorization method' allows us to study
interesting complex-action systems in various branches of physics.


\begin{acknowledgments}
We would like to thank J.\ Ambj\o rn, H.\ Aoki,
W.\ Bietenholz, Z.\ Burda, S.\ Iso, H.\ Kawai, E.\ Kiritsis, P.\ Orland,
M.\ Oshikawa, B.\ Petersson and G.\ Vernizzi for discussions.
The computation has been done partly on 
Fujitsu VPP700E at The Institute of Physical and Chemical Research (RIKEN),
and NEC SX5 at Research Center for Nuclear Physics (RCNP) of Osaka
University.
K.N.A.'s research was partially supported by RTN grants
HPRN-CT-2000-00122, HPRN-CT-2000-00131 and HPRN-CT-1999-00161 and the INTAS
contract
N 99 0590.
The work of J.N.\ was supported in part by Grant-in-Aid for 
Scientific Research (No.\ 14740163) from 
the Ministry of Education, Culture, Sports, Science and Technology.

\end{acknowledgments}

\bibliography{apssamp}

\begin{thebibliography}{99}
\bibitem{endnote0} 
It was found in the $D=4$ case that operators related 
to string theory possess well-defined large $N$
limit when $g^2 N$ is kept fixed \cite{AABHN}.
\end{thebibliography}

\end{document}